\documentclass{jaa}
\usepackage{natbib}
\usepackage{natbib}

\usepackage{graphicx}
\usepackage{url}
\usepackage{hyperref}
\hypersetup{
    colorlinks=true,
    linkcolor=red,
    filecolor=magenta,      
    urlcolor=magenta,
    citecolor=blue,
}

\begin{document}\sloppy

\title{A type II solar radio burst without a coronal mass ejection association}


\author{Anshu Kumari\textsuperscript{1} and Nat Gopalswamy\textsuperscript{2}}
\affilOne{\textsuperscript{1}Udaipur Solar Observatory, Physical Research Laboratory, Dewali, Badi Road, Udaipur-313 001, Rajasthan, India.\\}
\affilTwo{\textsuperscript{2}NASA Goddard Space Flight Center, 8800 Greenbelt Road Greenbelt, MD, 20771, USA.}


\twocolumn[{

\maketitle

\corres{anshu@prl.res.in}


\begin{abstract}
Type II solar radio bursts are commonly associated with shocks generated by coronal mass ejections (CMEs), where plasma waves are excited by magnetohydrodynamic (MHD) processes and converted into radio waves at the local plasma frequency or its harmonics. However, there are instances where type II bursts occur in the absence of whitelight CMEs. We analysed one such metric type II radio burst observed on November 2, 2023, characterized by split band features and fundamental-harmonic lanes. Notably, no CME was detected with space-based coronagraphs during this event. However, an intense M1.6 class flare was observed just before the type II burst and an extreme ultraviolet (EUV) disturbance was observed expanding into surrounding regions. The absence of any whitelight CME seen in any coronagraph field of view even though the EUV shock had a moderate speed of $\approx500~km/s$, which was close to the shock speed derived from radio observations, 
These observations indicate that the shock in the inner corona was most-likely driven by the EUV ejecta seen in the lower corona, but the ejecta did not survive as a CME in the coronagraph field of view.
\end{abstract}


}]


\doinum{12.3456/s78910-011-012-3}
\artcitid{\#\#\#\#}
\volnum{000}
\year{0000}
\pgrange{1--}
\setcounter{page}{1}
\lp{1}

\section{Introduction}
Shock waves are a fundamental phenomenon in astrophysics, capable of accelerating both electrons and ions ~\citep[e.g.][]{Thomsen1985, Mann1995}. In the solar atmosphere, shocks manifest as wavefront-like structures in multi-wavelength images or inferred from radio dynamic spectrum ~\citep[see for example,][]{Smerd1962, gopalswamy2013first}. Solar radio bursts, produced by energetic processes like solar flares and coronal mass ejections (CMEs), are classified based on their dynamic spectral characteristics into types I–V ~\citep{Wild1963}. Specifically, type II and IV bursts are linked with CMEs, indicating the critical role of CMEs in their production ~\citep[for example][and the references therein]{ Dulk1980, Morosan2021, Kumari2021}. In contrast, type III bursts are often associated with flares or other solar activities, representing electron beams traveling along open magnetic field lines.

Type II radio bursts are particularly noteworthy due to their association with electron beams accelerated by CME-driven shock waves \citep[e.g.][]{Smerd1975, Gopalswamy2005, kumari2017b}. These bursts are characterized by slowly drifting emission lanes observed at meter to decameter-hectometer (DH) and kilometer  wavelengths, usually at the plasma frequency's fundamental and/or harmonic \citep{Smerd1975}. These emission lanes frequently exhibit split bands and numerous fine structures \citep[see for example,][ and the references therein]{Smerd1975, Vrsnak2001, Morosan2022b, 2025A&A...697L...9K}. Generally, type II bursts and their fine structures are considered as signatures of expanding CME-driven shocks in an inhomogeneous solar corona \citep[e.g.]{ca87, Anshu2017a, Carley2020}. Traditionally, type II radio bursts are linked to shocks generated by CMEs in the solar corona. However, \cite{Kumari2023b} reported that less than $\approx4\%$ of type II radio bursts during the previous solar cycle lacked clear temporal and spatial associations with white-light CMEs.

\begin{figure*}[ht!]
    \centering
  \includegraphics[width=0.6\textwidth]{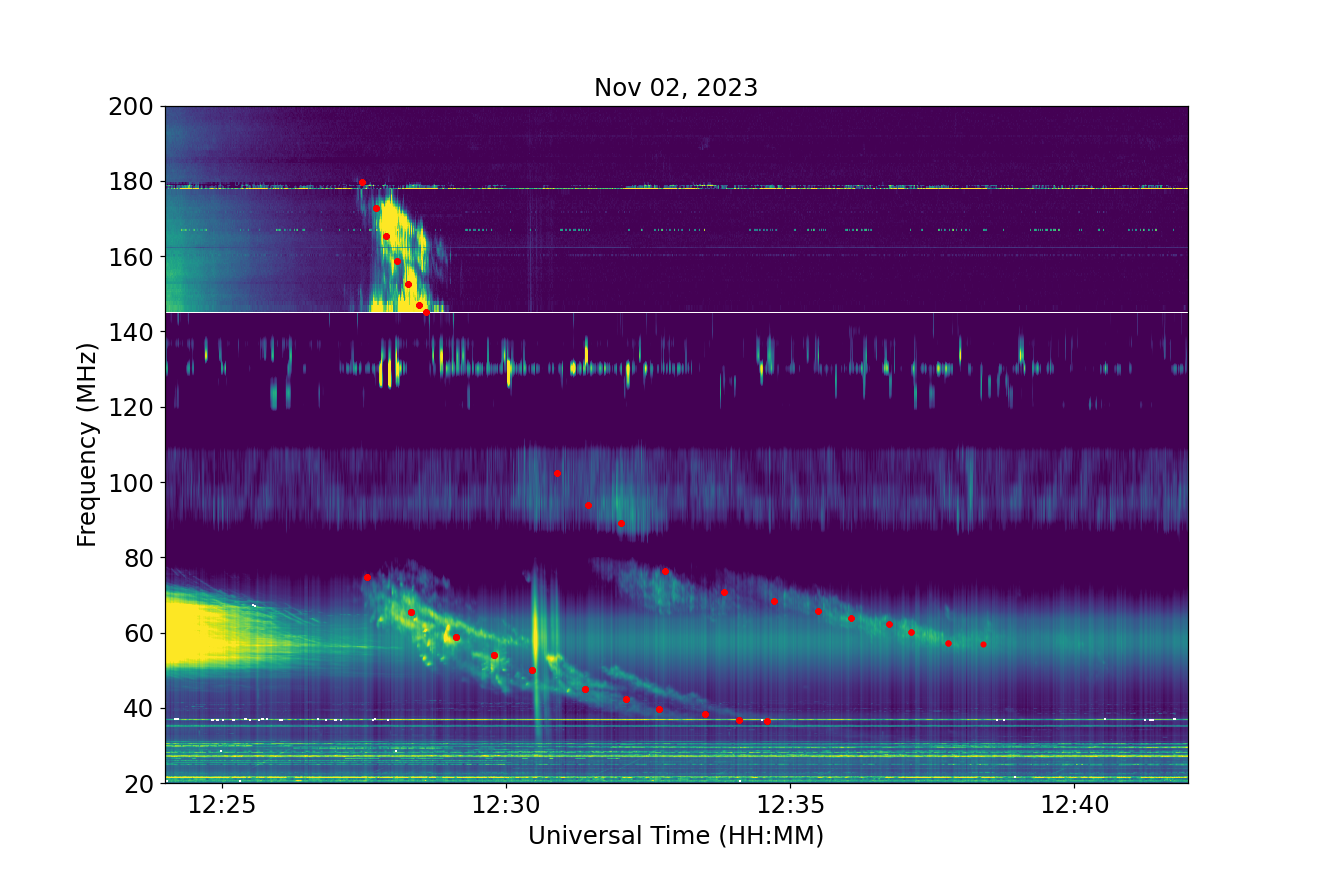}
  \vspace{-0.3cm}
  \caption{Dynamic spectra of the type II solar radio burst on 02 Nov 2023. The lower band is with the LOFAR data (20-80 MHz), the middle band is with e-CALLISTO (Birr), and the upper band is with ORFEES observations. The fundamental and harmonic bands are marked with dots in the spectra. We did the radio imaging for the harmonic part of the type II burst (see Fig.~\ref{fig1:figure2}).}
  \label{fig1:figure1}
\end{figure*}

Recent advancements in remote sensing high spatial and temporal resolution space solar instruments, like the \textit{Atmospheric Imaging Assembly (AIA)} \citep{lemen2011atmospheric} on board the \textit{Solar Dynamics Observatory} (SDO), the \textit{Extreme Ultraviolet Imager (EUVI)} \citep{Wuelser2004} on board the \textit{Solar TErrestrial RElations Observatory (STEREO)}, the \textit{Extreme Ultraviolet Imager (EUI)} \citep{Rochus2020} on board the \textit{Solar Orbiter  ((SolO)}, and the newly launched the \textit{Solar Ultraviolet Imaging Telescope (SUIT)} \citep{Tripathi2017} on board the \textit{Aditya-L1} have significantly improved our ability to observe and understand these phenomena by providing high-resolution images of the low corona, the formation heights of type II solar radio bursts \citep{Ma2011, Gopalswamy2012}.

Despite the consensus that CMEs are the primary source of metric type II radio bursts, some high-frequency type II bursts ($>150~ MHz$) have been suggested to originate from flare-generated shocks due to their speeds being fast and the absence of any white-light CME \citep{magdalenic2012flare, su2015, kumar2016flare, Morosan2023b}.
Some low-frequency type II bursts ($< 100~ MHz$) are caused by piston-driven shock waves from active region jets rather than CMEs \citep{Maguire2021}. Nonetheless, statistical studies have shown that most type II bursts are associated with CMEs, with only a small percentage lacking CME association \citep{Morosan2021, Kumari2023b}. An earlier investigation found that metric type II bursts without white-light CME signatures all had EUV signatures and originated from close to the disk center of the Sun \citep{2001JGR...10629219G}, the CMEs may not have been observed due to visibility issues \citep{2005JGRA..11012S05Y}.

In this article, we present a type II solar radio burst which had no white-light CME association as seen in coronagraphs. This article is organized as follows: in section \ref{section:sec2} we present the observations and data analysis, in section \ref{section:sec3} we present results and discuss the results and conclude the article in section \ref{section:sec4}. 

\begin{table*}[]
\caption{Event timeline in different wavelengths}
\centering
\label{tab:my-table}
\begin{tabular}{|c|c|c|l|l|}
\hline
                 & Type II    & X-ray        & EUV      & Whitelight CME \\ \hline
Start Time (UT)  & 12:27:21 & 12:21:30    & 12:23:30 & No  \\ \hline
End Time (UT)    & 12:38:09 & 12:26:10    & 12:36:00 & No  \\ \hline
Start Freq (MHz) & 180      & --          & --       & --  \\ \hline
End Freq (MHz)   & 36       & --          & --       & --  \\ \hline
\end{tabular}
\end{table*}

\section{Observations and data analysis}

\begin{figure*}[t]
    \centering
  \includegraphics[width=1\textwidth]{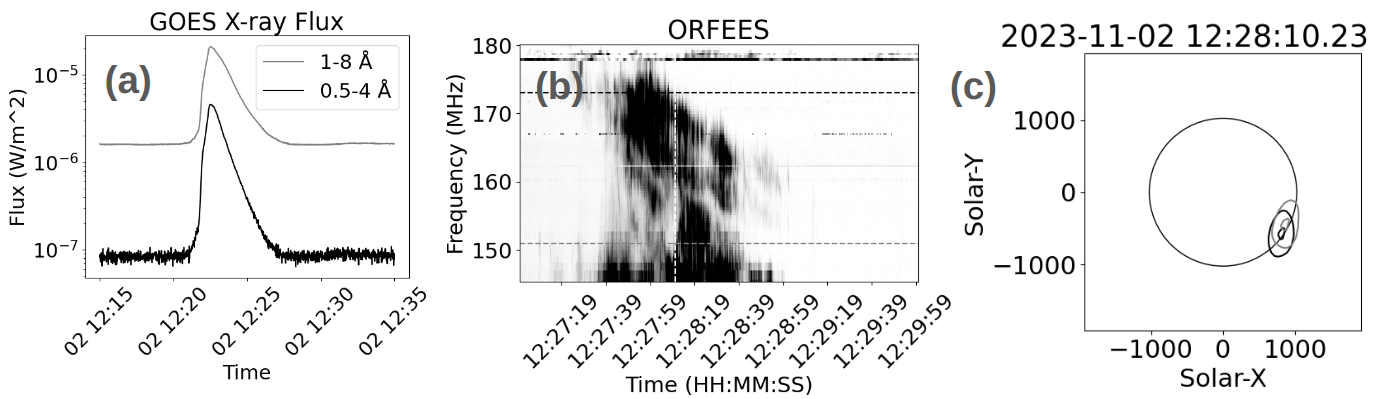}
  \vspace{-0.3cm}
  \caption{(a): The X-ray flare as seen with the GOES-16 satellite. This was a M1.6 class flare with a very short duration of around 4 minutes and 40 seconds, as shown in the figure. (b): Zoomed-in dynamic spectrum of the type II solar radio burst. The vertical dashed line shows the time of the image made in panel (c). The two horizontal lines show the frequency of the radio images in panel (c). (c): The radio contours (at $70\%$ level) at  show the location of the type II radio bursts at 170 MHz ('grey' color) and 150.9 MHz ('black' color) with the data obtained from the NRH instrument.
}
  \label{fig1:figure2}
\end{figure*}

A type II solar radio burst was observed on November 2, 2023 by several instruments: from the Low Frequency Array (LOFAR; \citealt{Haarlem2013}), e-CALLISTO (located at Birr) (\citealt{zu12}), and the Observation Radio pour FEDOME et l'Étude des Éruptions Solaires (ORFEES; \citealt{orfees21}). 
The burst commenced at $\approx12:27$ UT, exhibiting both fundamental and harmonic emission lanes with band splitting as illustrated in Fig.~\ref{fig1:figure1} and detailed in the zoomed-in region in Fig.~\ref{fig1:figure2}. Concurrently, an M1.6 class solar flare was detected by the GOES-16 instrument (AR13474, S18W30), beginning at $\approx12:21$ UT, peaking at $\approx12:23$ UT, and lasting approximately 4 minutes and 40 seconds, with a rise time of about 1 minute and 30 seconds, notably shorter compared to typical M-class flares that generally last for about 10-20 minutes \citep{daSilva2021} (see panel (a) of Fig.~\ref{fig1:figure2}). 
Extreme ultraviolet observations from both the SDO/AIA across all spectral channels and the EUVI on STEREO-A \citep{Howard2008} indicated brightening and faint moving structures. 
We use images from the Nan{\c c}ay Radioheliograph (NRH; \citealt{Kerdraon1997}), to locate the radio emission source in the solar corona. The NRH detected harmonic emissions of the type II burst across the frequencies 150.9 MHz and 173 MHz (see panel (c) of Fig.~\ref{fig1:figure2}). Radio imaging of the type II burst, with panel c showing two contour levels ($\approx 70\%$ of the peak intensity) for each frequency, representing the emission structure. The detection significance is approximately 5$\sigma$ above the noise level, where $\sigma$ = 0.1 SFU, measured as the noise in the solar radio map far from the Sun.

\begin{figure*}[t]
    \centering
  \includegraphics[width=0.38\textwidth]{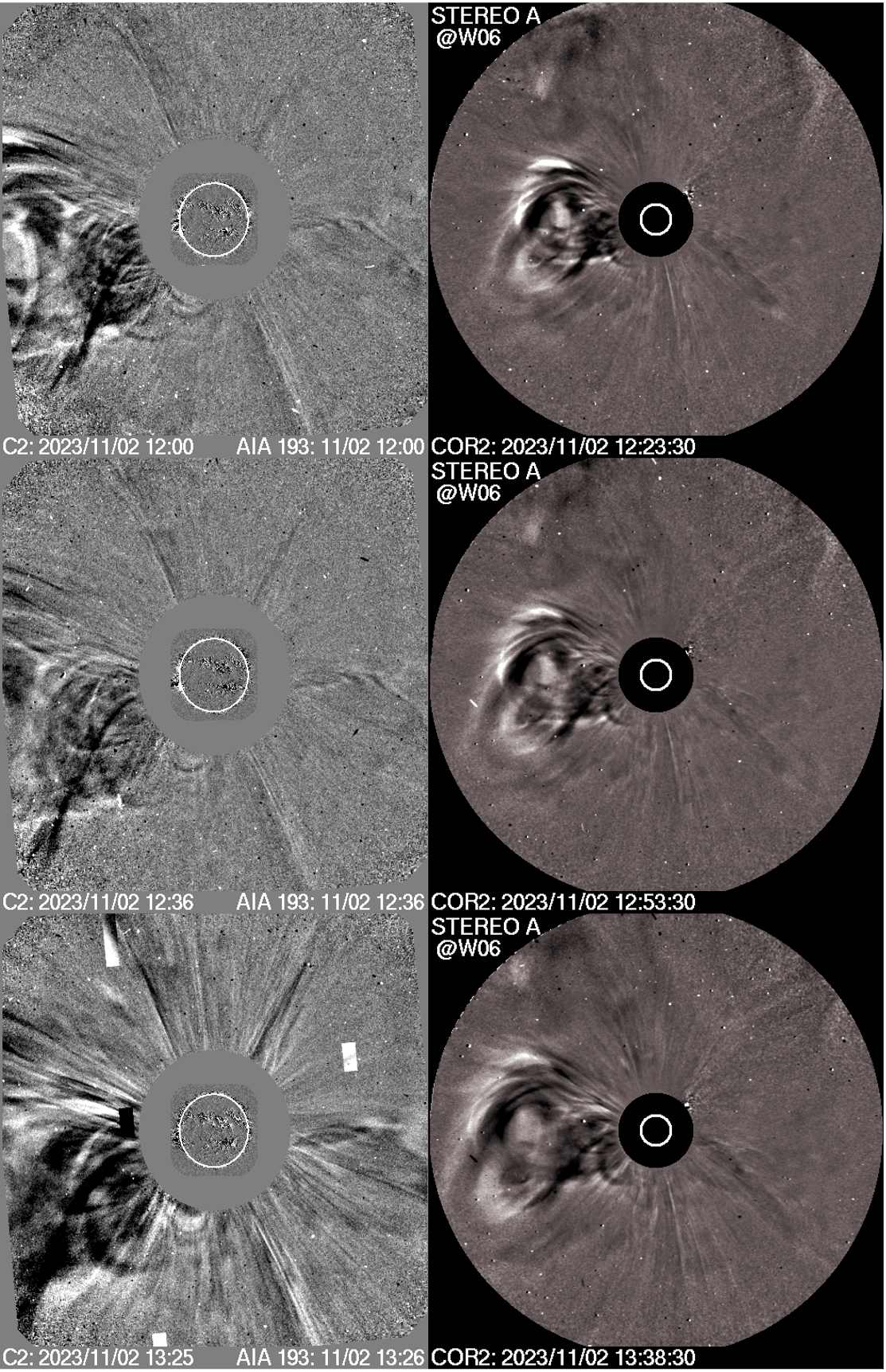}
  \includegraphics[width=0.595\textwidth]{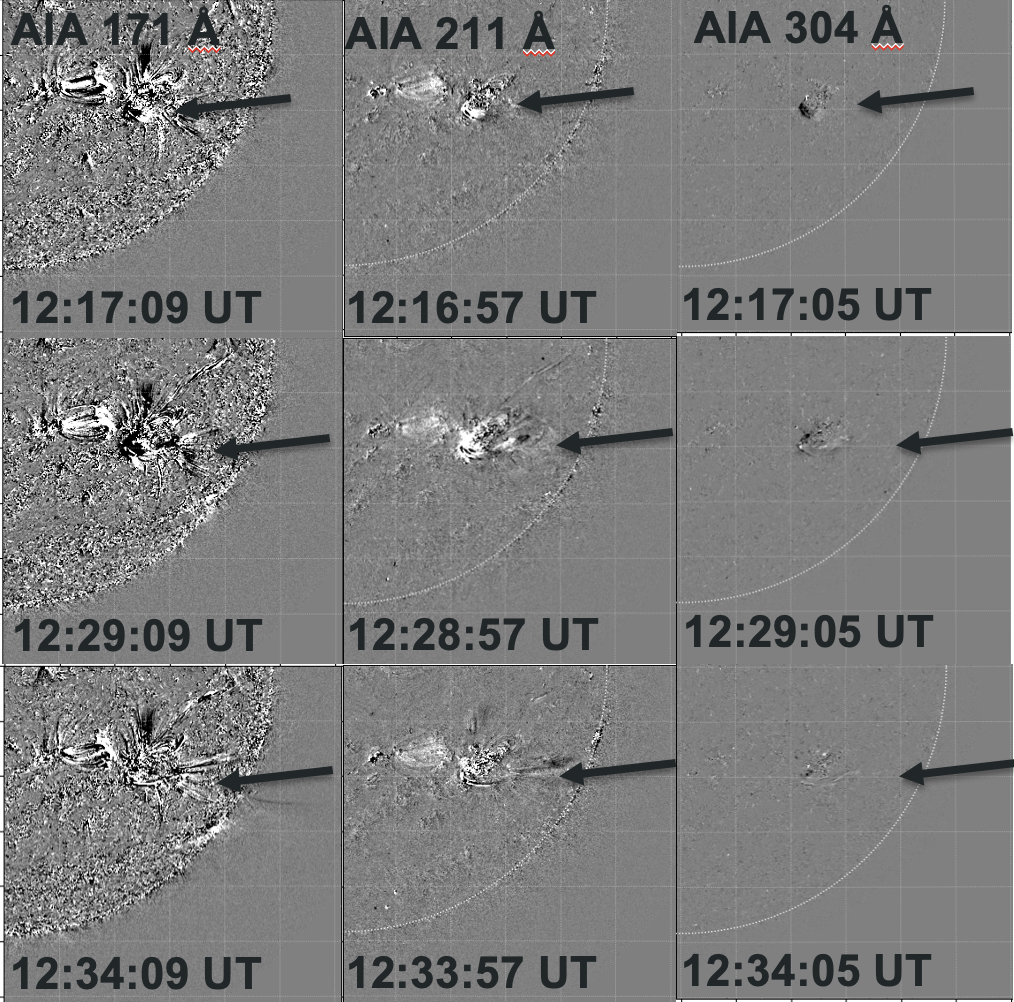}
  \vspace{-0.3cm}
  \caption{Left panel: Snapshots at different time instances from the two coronagraphs, SOHO/LASCO-C2 and STEREO-B/SECCHI-COR2 images. These show that no white-light CME was observed in the coronagraph FOV for the observed type II radio burst. Right Panel: Snapshots of different channels of SDO/AIA wavelengths images at different times. The arrow shows the movement of a faint feature in the AIA FOV.}
  \label{fig1:figure5}
\end{figure*}

Interestingly, no white-light CME (see Fig.~\ref{fig1:figure5}) was observed with the COR1 and COR2 coronagraph of the Sun-Earth Connection Coronal and Heliospheric Investigation (SECCHI; \citealt{Howard2008} on board the Solar Terrestrial Relations Observatory (STEREO) and the Large Angle and Spectrometric Coronagraph (LASCO) on the Solar and Heliospheric Observatory (SOHO; \citealt{Brueckner1995}). Therefore, this study investigates the origin of the type II solar radio burst. Table~\ref{tab:my-table} lists the details of the features observed in various wavelengths for this event. 

Fig.~\ref{fig1:figure5} shows snapshots taken by remote sensing instruments in white-light (left panel) and EUV wavelengths (right panel). Despite several CMEs observed in the coronagraph field of view (FOV), these events occurred significantly before or after the onset of the type II radio burst. During this period, STEREO was positioned approximately 6 degrees west of Earth, aligning LASCO and STEREO to observe the Sun from nearly the same perspective. No brightenings or disturbances were observed in the southwest region of the coronagraph FOV, confirming the absence of a white-light signature with this type II radio burst. However, immediately after the flare time (around 12:24 UT onward), a faint, narrow, and small moving disturbance was detected in the EUV FOV across all channels of SDO/AIA (left panel of Fig.~\ref{fig1:figure5}, channels 171 Å, 211 Å, and 304 Å). Thus, there is some mass motion associated with the eruption, even though there was no white light signature.

Fig.~\ref{fig1:figure1} illustrates the dynamic spectra of a type II solar radio burst, which occurred immediately following an M1.6 class solar flare. 
Remote observations were conducted using ground-based radio facilities, specifically the Nan{\c c}ay Radioheliograph (NRH; \citealt{Kerdraon1997}), to locate the radio emission source on the solar corona. The NRH detected harmonic emissions of the type II burst across the frequencies 150.9 MHz and 173 MHz (see panel (c) of Fig.~\ref{fig1:figure2}).

\label{section:sec2}

\section{Results and discussion}
\label{section:sec3}

The apparent motion of the source location of type II radio bursts along the line of sight (LOS) precluded the direct measurement of shock speed. Instead, the shock speed was estimated using the solar dynamic spectrum. 
Electron number densities at various frequencies (and hence heights) of the Sun can be derived from the type II burst frequencies observed in the dynamic spectrum. Using the plasma frequency and density relation, 
\begin{equation}
f_{p} [\text{MHz}] = 8.93 \times 10^{-3} \sqrt{N_{e} [\text{cm}^{-3}]}
\label{eq:plasma-frequency}
\end{equation}
where \( f_{p} \) is the plasma frequency in MHz and \( N_{e} \) is the electron number density in cm\(^{-3}\). Utilizing this density and Newkirk's density model \citep{newkirk1961solar} with a 4-fold enhancement factor \citep{Kumari2023b}, we estimated radio heights corresponding to the type II frequencies:
\begin{equation}
N_{e} [\text{cm}^{-3}] = E \times 4.2 \times 10^{4} \times 10^{4.32 \left( \frac{R_{\odot}}{r} \right)}
\label{eq:electron-density}
\end{equation}
where \( E \) is the density enhancement factor, \( R_{\odot} \) is the solar radius, and \( r \) is the height corresponding to the density estimated by Equation 1. We derived the shock speed from the drift rate using this relation \citep{2011pre7.conf..325G}:
\begin{equation}
V_{\text{shock}} = 2L \left( \frac{1}{f} \frac{df}{dt} \right)
\end{equation}
where \( L \) is the scale height derived by differentiating \( N_e \) in equation 2. We measure the width of the band splitting for the fundamental Type-II burst in the dynamic spectrum (see Fig. \ref{fig1:figure1}). The bandwidth is defined as
\begin{equation}
\text{BW} = \frac{F_u - F_l}{F_l}
\end{equation}
where \( F_u \) and \( F_l \) represent the upper and lower band frequencies, respectively. The density jump \([X]\) across the shock front is related to the bandwidth by
\begin{equation}
X = (\text{BW} + 1)^2
\end{equation}
This density jump \([X]\) is further connected to the Alfvénic Mach number \([M_A]\) through the equation:
\begin{equation}
M_A = \sqrt{\frac{X(X + 5)}{2(4 - X)}}
\end{equation}
This was used to calculate the average \(M_A\). With the shock speed obtained from equation 3, we derived the Alfvén speed. 
\begin{equation}
V_A = \frac{V_{\text{shock}}}{M_A}
\end{equation}

To estimate the split-band type II frequencies, which includes the upper and lower bands from in the combined dynamic spectra as shown Fig \ref{fig1:figure1}, we constructed frequency profile plots for each time step across the interval 12:28–12:35 UT. For each plot, the two prominent peaks corresponding to $F_u$ and $F_l$ were identified, and the center (peak) frequency of these profiles was taken as the estimated value. To quantify uncertainty, the frequency spread around each peak was used as an error bar. This method was validated by cross-referencing with faint e-CALLISTO spectra from Humain station\footnote{\url{https://soleil.i4ds.ch/solarradio/callistoQuicklooks/?date=20231102}}, which exhibit two clearer split-band lanes; for instance, at 12:33:30 UT, $F_u = 175$ MHz and $F_l = 150$ MHz, and at 12:34:30 UT, $F_u = 60$ MHz and $F_l = 125$ MHz. We note that the lower band in the LOFAR data is notably fragmented and less well-formed, deviating from a classic split-band structure, however, we used the aforementioned method for LOFAR spectral duration as well in the absence of any other observations. These values are provided in Table \ref{tab:my-table1}.

\begin{table*}[htbp]
\centering
\caption{Estimated parameters of the split-band Type II burst.}
\begin{tabular}{ccccccccc}
\hline
\textbf{Time (UT)} & \textbf{$F_u$ (MHz)} & \textbf{$F_l$ (MHz)} & \textbf{BW} & \textbf{X} & \textbf{$M_a$} & \textbf{$V_a$ (km/s)} \\
\hline
12:27:31 & 177   & 160.5  & 0.103 & 1.216 & 1.165 &  -- \\
12:27:35 & 176.5 & 159    & 0.110 & 1.232 & 1.178 & 489.4 \\
12:27:37 & 175.5 & 158    & 0.111 & 1.234 & 1.179 & 594.9 \\
12:27:45 & 173   & 156.5  & 0.105 & 1.222 & 1.170 & 513.3 \\
12:27:49 & 172.5 & 155.5  & 0.109 & 1.231 & 1.177 & 408.5 \\
12:27:54 & 171   & 154   & 0.110 & 1.233 & 1.178 & 495.7 \\
12:27:59 & 170   & 153    & 0.111 & 1.235 & 1.180 & 334.4 \\
12:28:04 & 168   & 152   & 0.105 & 1.222 & 1.170 & 340.8 \\
12:28:09 & 167.5 & 151   & 0.109 & 1.230 & 1.176 & 342.4 \\
12:28:19 & 165   & 147.5 & 0.119 & 1.251 & 1.193 & 505.2 \\
12:28:29 & 161   & 143   & 0.126 & 1.268 & 1.206 & 504.1 \\
12:33:35 & 80    & 76    & 0.053 & 1.108 & 1.082 & 545.5 \\
12:34:05 & 77    & 62    & 0.242 & 1.542 & 1.433 & 422.6 \\
\hline
\end{tabular}
\label{tab:my-table1}
\end{table*}


\cite{su2015} used a restricted density model to determine the shock speed responsible for the observed frequency drift rate in the dynamic spectrum. The inverted shock speed closely matched the speed of the wavefront-like structure, indicating that the structure is likely a coronal shock generating the type II radio burst. Similarly, we investigated the EUV disturbance observed in SDO/AIA channels (see right panel of Fig.~\ref{fig1:figure5}) and derived its speed, comparing it with the speed obtained from the dynamic spectrum. The shock speed derived from both methods showed consistency with the EUV disturbances. A similar event was identified by \cite{Morosan2023b}, where a band-split type II radio burst associated with a C-class solar flare was reported without any concurrent detection of a CME. The authors suggested that the shock wave observed could have been generated either by bulk plasma motions associated with a failed eruption or the flaring process itself. In our study, akin to theirs, we investigated the propagation of an EUV disturbance within the SDO/AIA FOV. The EUV front propagated initially at a speed of approximately 500 km/s and is anticipated to intensify into a shock wave within a region (approximately $1.5~R_{\odot}$) characterized by a low Alfv\'en speed.

\begin{figure*}[t]
    \centering
  \includegraphics[width=0.45\textwidth]{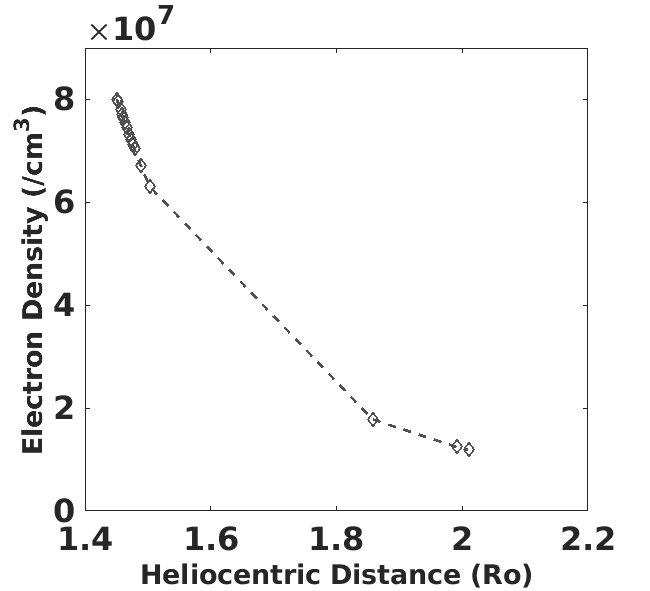}
  \includegraphics[width=0.45\textwidth]{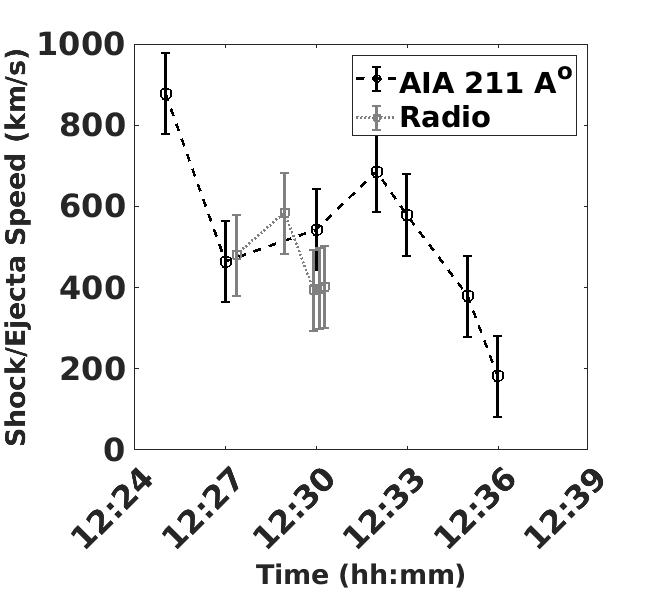}
  \vspace{-0.3cm}
  \caption{The electron density obtained from radio observations using equation 2 (left panel). The ejecta speed plot showing the speed of the disturbance in the EUV field of view and the shock speed derived from the radio observations (right panel).}
  \label{fig1:figure4}
\end{figure*}

An important point to highlight here is that around 150 MHz, corresponding to approximately 1.45-1.5 $~R_{\odot}$ in height in the corona, the Alfv\'en speed was found to be approximately 350 km/s. This condition is ideal for producing a type II radio burst, as even a relatively weak shock speed is sufficient under these circumstances \citep{2001JGR...10629219G, mann2003formation}. 
\cite{2001JGR...10629219G} has shown that type II bursts without CMEs preferentially originated from the disk center of the Sun. The authors also found that 11 out of 34 metric Type II bursts from source longitudes less than \(60^\circ\) were not linked to LASCO CMEs, suggesting that about one-third of the CMEs originating from the disk center were not detectable by LASCO. We note that the disk center location reduces the visibility of white light CMEs \citep{2005JGRA..11012S05Y}. \cite{Gopalswamy2012} has shown that, the EUV counterpart of a CME can have a higher speed than the white-light CME. This means the CME can be driving a shock close to the Sun but becomes too weak to drive a shock by the time it reaches the coronagraph field of view. The present case is an extreme version of this scenario in that the CME became too weak to be observed. We note that the speed variation in Fig. \ref{fig1:figure4} is very similar to the ones reported by \cite{gopalswamy2009euv, Gopalswamy2012}. The error bars for the radio speeds, and AIA speeds were calculated using the bootstrap method, involving ten resampled datasets from the frequency drift measurements over the interval 12:28–12:35 UT.
\citet{Gopalswamy2009} presented evidence of a similar short-duration flare that was associated with both a CME and a type II radio burst. However, it is important to note that the event was located near the limb of the Sun. Limb events are generally easier to detect in white light, as CMEs projected against the solar disk (which has origin on the disc) can become difficult to distinguish due to the line-of-sight integration effects. 
The current work is based on the widely accepted framework that the upper $(F_u)$ and lower $(F_l)$ split bands of the type II burst (as visible from Fig \ref{fig1:figure1} originate from forward and reverse shocks propagating through the solar corona, a model rooted in the interpretation of band-splitting as a manifestation of shock dynamics in magnetized plasma \citep{Smerd1975}. This assumption leverages the differential propagation speeds of forward and reverse shocks, which modulate the local plasma frequency and generate the observed frequency separation between $F_u$ and $F_l$. However, alternative scenarios exist, such as the bands arising from different regions of a single shock \citep{1967PASA....1...47M, 2011sswh.book..267C}. More recently, with advanced radio imaging instruments, studies have shown that these emission can come from spatially distinct emission sites \citep{2023A&A...670A.169B, 2025SoPh..300..108F}. These possibilities, including multi-band structures not explained by the forward/reverse model, should be noted for future investigation. A notable discontinuity in the drift rate of the type II emission is observed just before $\approx 12:31$ UT in the LOFAR data (see Fig. 1), potentially reflecting a change in shock geometry or a secondary acceleration event, which needs to be further investigated with radio imaging observations.

\section{Conclusions}
\label{section:sec4}

We investigated a relatively rare occurrence of type II solar radio bursts not associated with aEUV disturbance seen in the lower corona. The EUV ejecta had a moderate speed of white-light CME. The type II burst occurred on November 2, 2023, coinciding with an intense M1.6 class flare and exhibited fast (\(\sim500 \, \text{km/s}\)) EUV disturbances. Given that the eruption originated near the disk center, it is plausible that these EUV disturbances represent the near-Sun manifestation of a CME. Though we note that, due to visibility issues inherent in observing disk-centered events, the CME likely escaped detection within the FOV of the coronagraph. This scenario is consistent with the findings of \citet{2001JGR...10629219G} and \citet{2005JGRA..11012S05Y}, who showed that CMEs originating near the solar disk center are often not observed by coronagraphs due to projection effects and reduced contrast against the bright solar disk. 
The key findings from our study include:

\begin{enumerate}
\item The starting frequency of type II burst (180 MHz [harmonic]) corresponds to the  minimum in the Alfv\'en speed ($\approx 350~km/s$) profile.

   
\item The shock was inferred from an EUV disturbance, which is likely to be the CME manifestation near the Sun, even though there was no white-light manifestation.
   
\item No coronal mass ejection was observed in the fields of view of both SOHO/LASCO-C2 and STEREO-A/SECCHI-COR1 and COR2. However, the faint, moving feature observed in the EUV FOV was associated with the type II radio burst is likely to be the near-Sun manifestation of the CME. Disk-center eruptions associated with type II burst and EUV disturbance are known to lack white-light CMEs due to visibility issues \citep{2001JGR...10629219G,2005JGRA..11012S05Y}.
   
\end{enumerate}

This observation of a type II burst without a CME adds to the limited catalog of such events, providing a valuable dataset for refining models of coronal shock propagation and radio emission mechanisms. The lack of a visible CME association suggests alternative drivers, such as flare blast wave \citep{magdalenic2012flare}, a failed eruption \citep{Morosan2023b}, CMEs below the detection threshold \citep{2005JGRA..11012S05Y}, which needs further investigation with instruments like Visible Emission Line Coronagraph (VELC) onboard ADITYA-L1 for detection in low corona\citep{2024ApJ...976L...6R}.

\section*{Acknowledgments}
AK acknowledges the ANRF Prime Minister Early Career Research Grant (PM ECRG) program. AK was supported by the NASA Postdoctoral Program at the the NASA Goddard Space Flight Center (GSFC). NG is supported by NASA's Living With a Star program and the STEREO project. We thank the SDO, SOHO, STEREO teams for making the data openly available. We also acknowledge the e-CALLISTO network and the IDOLS (Incremental Development of LOFAR SpaceWeather) project for radio spectral data and the the Nan{\c c}ay Radioheliograph for imaging data.  
\vspace{-1em}

\bibliography{reference}
\bibliographystyle{apj}

\end{document}